\documentclass[12p]{article}

\usepackage{mathrsfs}
\usepackage{epstopdf}
\usepackage[caption=false]{subfig}
\usepackage[numbers,sort&compress]{natbib}
\usepackage{amsmath}
\usepackage{graphicx}
\usepackage[affil-it]{authblk}

\begin{document}

\title{A Chi-square Goodness-of-Fit Test for Continuous Distributions against a known Alternative.}

\author{Dr. Wolfgang Rolke}

\author{Cristian Gutierrez Gongora}

\affil{Dept. of Mathematical Sciences, University of Puerto Rico - Mayaguez}

\maketitle

\begin{abstract}
The chi square goodness-of-fit test is among the oldest known
statistical tests, first proposed by Pearson in 1900 for the multinomial
distribution. It has been in use in many fields ever since. However,
various studies have shown that when applied to data from a continuous distribution it is
generally inferior to other methods such as the Kolmogorov-Smirnov or Anderson-Darling tests.
However, the performance, that is the power, of the chi square test depends
crucially on the way the data is binned. In this paper we describe a
method that automatically finds a binning that is very good against a
specific alternative. We show that then the chi square test is generally
competitive and sometimes even superior to other standard tests.

{\bf Keywords:} Kolmogorov-Smirnov, Anderson-Darling, Zhang tests, Power, Monte Carlo Simulation
\end{abstract}

\section{Introduction}

A goodness-of-fit test is concerned with the question whether a data set
may have been generated by a certain distribution. It has a null
hypothesis of the form \(H_0: F=F_0\), where \(F_0\) is a probability
distribution. For example, one might wish to test whether a data set
comes from a standard normal distribution. An obvious and often more
useful extension is to test \(H_0: F \in \mathscr{F}_0\) where \(\mathscr{F}_0\) is a family of
distributions but without specifying the parameters. So one might wish
to test whether a data set comes from a normal distribution but without
specifying the mean and standard deviation. If the null distribution is
fully specified we have a simple hypothesis whereas if parameters are to
be estimated it is called a composite hypothesis.

As described above a goodness-of-fit test is a hypothesis test in the
Fisherian sense of testing whether the data is in agreement with a
model. The main issue with this approach is that it does not allow us to
decide which of two testing methods is better. To solve this problem
Neyman and Pearson in the 1930s introduced the concept of an alternative
hypothesis, and most tests done today follow more closely the
Neyman-Pearson description, although they often are a hybrid of both.
The original Fisherian test survives mostly in the goodness-of-fit
problem, because here the obvious alternative is \(H_a: F \not\in \mathscr{F}_0\), a
space so huge as to be useless for power calculations.

However, in almost all discussions of goodness-of-fit testing there is
an immediate pivot and a specific alternative is introduced. This is a
necessary step in order to be able to say anything regarding the
performance of a test. In our method we have taken advantage of this,
and describe a way to find a good binning from among a large number.
 We will show that if such a binning is used the
chi-square test can be quite competitive with other tests, and sometimes
even better. Moreover it fully automates the choice of bins.

The goodness-of-fit test is one of the oldest and most studied problems in
Statistics. For an introduction to Statistics and hypothesis testing in
general see Casella and Berger (2002) or Bickel and Doksum (2015). For discussions of the many
goodness-of-fit tests available see D'Agostini and Stephens (1986), 
Raynor et al. (2012), Zhang (2002) and  Thas (2012). Chi-square tests are the
subject of Watson (1958), Greenwood and Nikulin (1996) and Voinov and Nikulin (2013). 
Thas (2012) has an extensive list of references on the subject.

\section{Chi-square test}

The original test by Pearson was designed to see whether an observed set
of counts \(O\) was in agreement with a multinomial distribution with
parameters \(m, p_1,..,p_k\). This is done by calculating the expected
counts \(E_i=m p_i\) and the test statistic \(X=\sum (O-E)^2/E\). Pearson
showed that under mild conditions \(X \sim \chi^2(k-1)\), a chi-square
distribution with \(k-1\) degrees of freedom. The test therefore rejects
the null at the \(\alpha\) level if \(X\) is larger than the
\((1-\alpha)100\%\) quantile of said chi-square distribution. Later work
showed that this test works well as long as the expected counts are not
to small. \(E>5\) is often suggested although it has been shown that it
can still work well even if some expected counts are smaller. In the
context of goodness-of-fit testing for a continuous distribution it is
generally possible to insure \(E>5\) for all bins, and we will do so in
all that follows.

If the test is to be applied to a continuous distribution this
distribution has to be discretized by defining a set of bins. In
principle any binning (subject to \(E>5\)) will work. Two standard
methods often used are bins of equal size and bins with equal
probabilities under the null hypothesis.

Among Statisticians the use of the chi-square test for continuous
distributions has been discouraged for a long time. This is due to its
lack of power when compared to other tests. However, it is still the
go-to test in many applied fields, and this will be the case for a long
time to come. The chi-square test does in fact have one advantage over 
most other test, namely that it deals with parameter estimation very
easily. Fisher (1922) showed that if the parameters are
estimated by minimizing the chi-square statistic, \(X\) again has a
chi-square distribution, now with \(k-1-p\) degrees of freedom where
\(p\) is the number of parameters estimated. In fact Fisher coined the
term degrees of freedom in this seminal paper. 

By contrast other standard tests do not easily generalize to allow
parameter estimation, and usually the null distribution has to be found
via simulation.

A number of alternative test statistics have been proposed, all of which
also lead to a limiting chi-square distribution. For a survey of such
statistics see D'Agostini and Stephens (1986) and  Cressie and Read (1989). In this paper we use six different
formulas and again find the one that yields the highest power against a specific alternative.
They are:

$$
\begin{aligned}
\text{Pearson: }   &\sum \frac{(O-E)^2}{E} \\
\text{Freeman-Tukey: } &4\sum (\sqrt O-\sqrt E)^2 ) \\
\lambda_p \text{: }  &2\sum(E-O+O\log(O/E))\\
\text{loglikelihood G}^2 \text{: } &2\sum(O\log(O/E)) \\
\text{Neyman Modified: } &\sum(E^2/O-O) \\
\text{lambda2/3: } &\frac{9}{5}\sum (O(O/E)^{2/3}-1)
\end{aligned}
$$

Many of these are members of the Cressie-Read divergence family (Cressie and Read 1989), given by

\begin{equation}
I^\lambda(q:p)=\frac1{\lambda(\lambda+1)} \sum q\left[(q/p)^\lambda-1 \right]
\end{equation}

\section{Parameter estimation}

If the test has a composite null hypothesis it is necessary to estimate
the parameters from the data. In practice this is often done via maximum
likelihood estimation using the unbinned data. That this leads to a test
that is anti-conservative was shown long ago by Fisher (1922). As
an example consider the following case: We wish to test whether the data
comes from a normal mixture, that is

\begin{equation}
\mathscr{F}_0=\lambda N(\mu_1, \sigma_1)+(1-\lambda) N(\mu_2, \sigma_2)
\end{equation}

and the parameters \(\lambda, \mu_1,\sigma_1, \mu_2, \sigma_2\) are to
be estimated. For our simulation we will generate \(1000\) observations
with \(\lambda =\frac13, \mu_1=0,\sigma_1=1, \mu_2=5, \sigma_2=2\) and
use unbinned maximum likelihood to estimate the parameters. Then we apply the
chi-square test with 10 equal probability bins at the \(5\%\) level.
This is repeated 10000 times. We find a true type I error rate of over
\(9\%\), almost double the nominal one of \(5\%\).

This simulation as well as all others discussed in this paper where done
using R. An Rmarkdown file with all calculations as well as an R library
with all routines is available at 
https://github.com/wolfgangrolke/chisqalt.

Instead of maximum likelihood with the unbinned data one can use either
maximum likelihood with binned data or a method called minimum
chi-square. This is just what the name suggests, find the set of values
of the parameters that minimizes the chi-square statistic. Using this
estimation method in the above simulation yields a correct type I error
rate of \(5\%\). The same is true in every simulation study we
performed, and our routine uses this estimation method. 

This method of estimation has another advantage: by the way it is
defined it is clear that if the null hypothesis is rejected, it would
also be rejected for any other set of parameter values. Also, Berkson (1980) 
argues in favor of minimum chi-square.

Watson (1958) discusses this issue and provides alternative formulas for the chi-square statistic when unbinned 
maximum likelihood estimation is used. This was useful when minimum chisquare was 
difficult to do in the absence of computers but is no longer necessary today.

\section{Other tests}

Many other goodness-of-fit methods have been developed over time. The
most commonly used are the Kolmogorov-Smirnov (KS) and Anderson-Darling
(AD) tests. Zhang (2002) proposed three statistics (ZK, ZA and ZC) based on likelihood ratios and
showed that they provided good power in a number of cases. We will employ all of these 
methods for comparison.  For the cases where
parameters need to be estimated we will use unbinned maximum likelihood
estimation and simulation to find the null distributions of these tests.
For discussions of the relative merits of these tests see  Massey (1951),
Birnbaum (1952), Goodman (1954), Zhang (2002) and Thas (2010).

\section{Binning}

\subsection{Bin types}

There are two binning methods that are most commonly used. They are bins of equal size
and bins with equal probabilities. In the case of equal size bins often
some adjustment is necessary to insure that all expected counts are at
least 5.

We will increase the bin types as follows. Let us say that it was decided to 
use k bins, and the equal probability
method yielded bins with endpoints \(B_i^0\) whereas the equal size bins
have endpoints \(B_i^1\), \(i=0,..,k\). Here possibly
\(B_0^0= B_0^1=-\infty\) and \(B_{k}^0= B_{k}^1=\infty\). We define a
new set of bins by interpolating between these, so a \(\kappa\) bin set
has endpoints

\begin{equation}
B_i^\kappa=(1-\kappa) B_i^0+\kappa B_i^1
\end{equation}

for any \(0\le \kappa \le 1\). Note that this includes equal probability
bins (\(\kappa=0\)) as well as equal size bins (\(\kappa=1\)).

\subsection{Number of bins}

Many studies in the past have focused on the number of bins to use. A
default formula used in many software programs, including R's hist function, 
is Sturges' rule \(k=1+\log_2(n)\), Sturges (1926),
where n is the sample size. One of the first suggestions was Mann and Wald (1942)
\(k=4\left[ \frac{2(n-1)^2}{c^2}]\right]^\frac15\).
Others can be found in D'Agostini and Stephens (1986),  Dahiya and Gurland (1973),
Bogdan (1995), Harrison (1985), Kallenberg (1985), Kallenberg et al (1985), 
Koehler and Gann  (1990), Williams (1950), Oosterhoff (1985) and Quine and Robinson (1985).
Mineo (1979) suggests a different binning scheme. 

\begin{figure}
  \includegraphics[width=\linewidth]{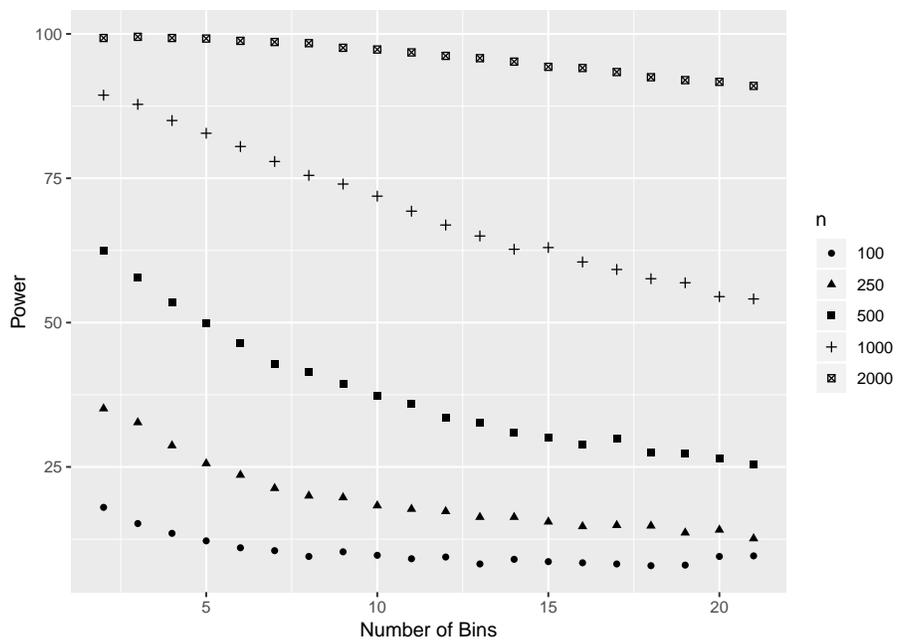}
  \caption{Power of testing uniform vs. linear. While power increases as sample size increases,
in each case just two bins yield the highest power.} 
  \label{figure1}
\end{figure}

All of these suggestions have in common that
that the number of bins increases as the sample size increases, although some  
recommend a small number (6 or 7) independent of sample size.
However, a simple example shows that the optimal number of bins does not always increase 
with the sample size. Say we wish
to test \(H_0: F=U[0, 1]\) vs \(H_1: F=\) Linear(slope=0.2). Here under
the alternative the distribution function is \(F(x)=0.2x^2+0.8x\text{; }0<x<1\). In this
case equal size and equal probability bins are the same. We run
simulations for the cases of 2 to 21 bins and sample sizes 100, 250, 500 
100 and 2000. The resulting powers are shown in figure~\ref{figure1}. The power
increases as the sample size increases but in each case just two bins
yields the highest power.

\subsection{Finding the optimal bin set}

Optimal in the context of hypothesis testing always means having the
highest power, and the power of a test can only be found when an
alternative is specified. So let us consider the following problem: we
wish to test \(H_0: F=F_0\) vs \(H_1: F=F_1\). A standard way to estimate the power would
proceed as follows: generate a data set of the desired size from
\(F_1\), apply the test to the data and see whether it rejects the null
at the desired level. Repeat many times, and the percentage of
rejections is the power of the test.

In the case of a chi-square test this means one has to find the bin
counts for the generated data set. However, we already know that these
counts have a multinomial distribution with probabilities
\(p_i=F_1(B_{i+1}^\kappa)-F_1(B_i^\kappa)\). We can therefore run the
simulation by generating variates from a multinomial distribution with
these probabilities directly. In our simulation studies we have seen a
speedup on the order of 100 and more using this approach. An additional
advantage is that we need not be able to generate variates from any
\(F_1\) directly. This is done in the cases where no parameter estimation is needed.

There is another issue when trying to find an optimal binning. How are
we to choose between two binnings that both have a power of 100\%? Moreover,
using (say) k and k+1 bins will often result in tests with almost equal
power, well within simulation error. To always find a single best
binning (from among our sets) we will use the idea of a \emph{perfect}
data set. This is an artificial data set that has its observations at
the exact right spots, under the alternative. These of course are simply
the quantiles of the distribution under the alternative hypothesis.
Applying the test to this perfect data set and using k, \(\kappa\) bins
we will use as a figure of merit
\(M_{k, \kappa, TS}=TS/{\text{qchisq}(0.95, k-1-p)}\), where \(TS\) is the
value of one of the test statistics discussed above and \({\text{qchisq}(0.95, k-1-p)}\) is the
\(95\%\) critical value of a chi-square distribution with k-1-p degrees of freedom. \(95\%\) was chosen here
not because we wish to test at the \(5\%\) level but to account
(roughly) for the increase in the critical values.

The idea here is simple: the binning that yields the highest value of
\(M_{k, \kappa, TS}\) would be best in rejecting the null hypothesis for the
perfect data set at the \(95\%\) level, and one would expect this to be
quite good for testing a random data set from that same distribution.

Let us apply this idea to the following case: we wish to test
\(H_0: F= \text{Linear}(\text{slope} = -0.5)\) vs
\(H_a: F= \text{ truncExp}(1)\), an exponential rate 1 truncated to
{[}0, 1{]}.

\begin{figure}
  \includegraphics[width=\linewidth]{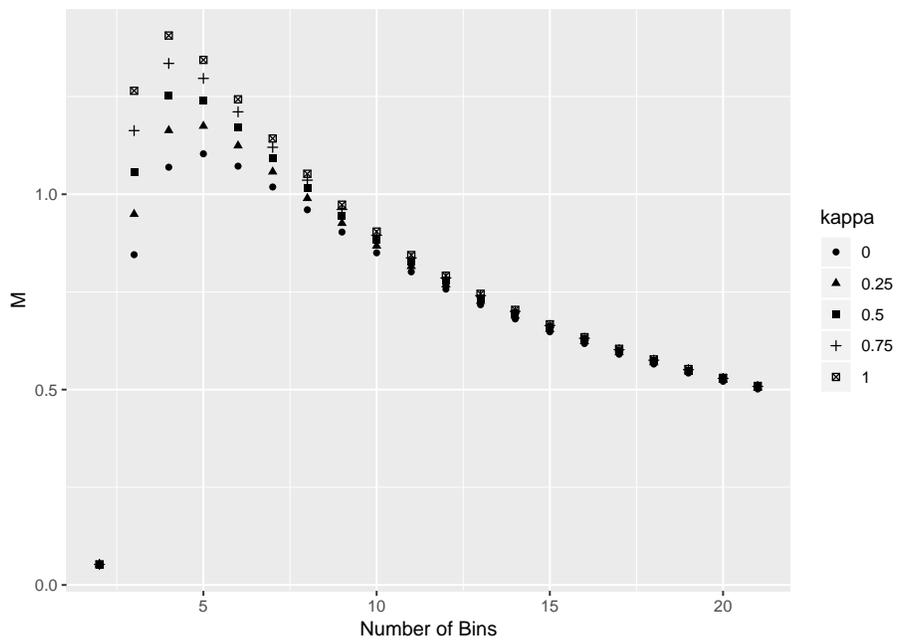}
  \caption{$M_{k, \kappa}$ of a test of linear vs truncated exponential. The maximum value of $M_{k, \kappa}$ is attained for $k=4, \kappa=1$}  
  \label{figure2}
\end{figure}

\begin{figure}
  \includegraphics[width=\linewidth]{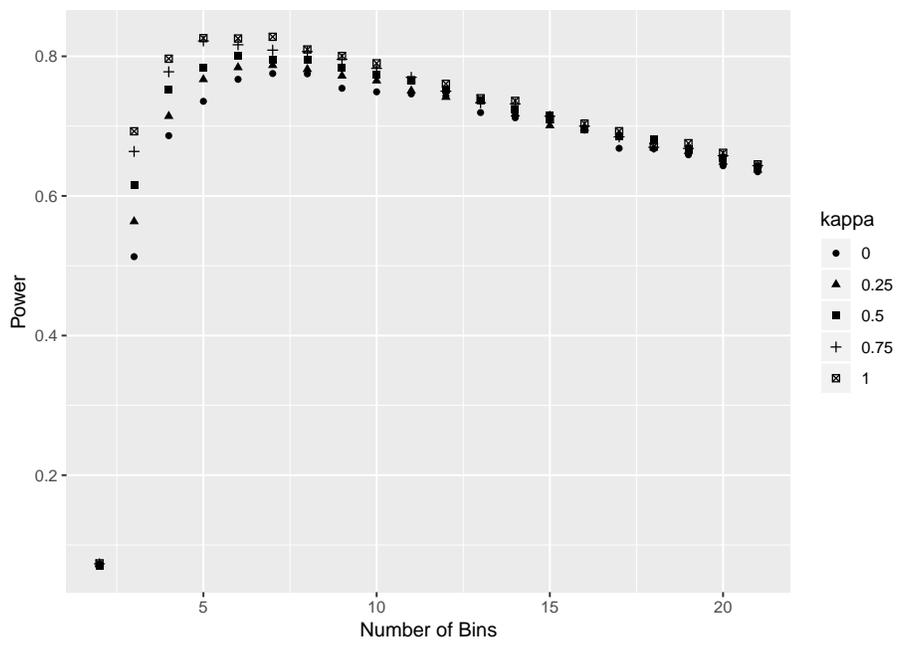}
  \caption{Power of a test of linear vs truncated exponential. The maximum power is attained for $k=4, \kappa=1$}  
  \label{figure3}
\end{figure}

Assuming a sample size of 10000, in figure~\ref{figure2} we have
\(M_{k,\kappa, TS}\) for \(k=2,..,21\), \(\kappa=0, 0.25, 0.5, 0.75, 1\)  and
using Pearson's formula. The highest value is achieved for \(k=4, \kappa=1\). In 
figure~\ref{figure3} we see the actual powers, and indeed (within simulation error)
\(k=4, \kappa=1\) is best.

So our test proceeds as follows: it searches through a grid of number
and type of bins as well as the formulas for the test statistics. By default these are \(k=(2+p):2(1+\log_2(n))\) and \(\kappa=0, 0.25, 0.5, 0.75, 1\),
but these can be adjusted by the user. It finds the combination that
maximizes \(M_{k,\kappa, TS}\) and applies the corresponding chi-square test
to the data. Therefore the choice of number of bins, type of
bins and test statistic is automatic, and independent of the data.

It is important to note that this searching through many binning schemes will not lead 
to a problem of multiple testing, and thereby a wrong type I error probability, because 
the test is done only once, using the chosen binning scheme.

As an example, if a data set has 1000 observations, the method searches through 660 different binning schemes and chi-square formulas. This only takes a fraction of a second on a modern PC.

\section{Other circumstances}

Our routines also allow for two situations sometimes encountered in
practice:

\subsection{Already binned data}

In some fields it is common that the data, although coming from a
continuous distribution, is already binned. This is typically the case,
for example, in high energy physics experiments because of finite
detector resolution. If so our routine finds the optimal binning as
described above and then finds the combination of the data bins that
comes closest to the optimal one.

\subsection{Random sample size}

Another feature often encountered is that the sample size itself is
random. This is the case, for example, if the determining factor was the
time over which an experiment was run. Our routine allows this if the
sample size is a variate from a Poisson distribution with known rate
\(\lambda\), as is often the case.

One consequence of such a random sample size is that the bin counts no
longer have a multinomial distribution, but instead are independent
Poisson with rates that depend on the bin probabilities and \(\lambda\).
In turn this implies that the chi-square statistic now has \(k-p\)
instead of \(k-p-1\) degrees of freedom. Another consequence is that for
the Kolmogorov-Smirnov, Anderson-Darling and Zhang tests the null
distribution has to be found via simulation, even if no parameters are
estimated.

\section{Performance}

\subsection{Type I error}

Our method always guarantees that the binning used has expected counts at least five, and so the
basic theory of chi-square tests should yield a good approximation to the chi-square distribution. 
Nevertheless, table 1 shows the results of a simulation study with a number of null distributions and
using $\alpha=1\%, 5\%$ and $10\%$. As it is based on 10000 simulation runs we see that the true type I error
probabilities are well within the simulation error of the nominal ones.

\begin{table}
\setlength\tabcolsep{12pt}
\begin{tabular}{l|c|c|c}
\hline\\[2pt]
Null Distribution & 1\% & 5\% & 10\%\\[2pt]
\hline\\[2pt]
U[0,1] & 1.07 & 4.94 & 9.95\\[2pt]
Beta(2,4) & 0.96 & 5.15 & 10.00\\[2pt]
Gamma(3, 0.5) & 0.88 & 5.09 & 9.78\\[2pt]
N(0,1) & 1.05 & 5.10 & 9.60\\[2pt]
Normal & 1.01 & 4.70 & 9.77\\[2pt]
Exp(1) & 1.07 & 4.76 & 9.93\\[2pt]
Exponential & 1.12 & 5.50 & 10.52\\[2pt]
\hline
\end{tabular}
\caption{True Type I error probabilities}
\end{table}

In the following sections we will discuss a number of cases. For each we will find
the power of the Kolmogorov-Smirnov (KS), Anderson-Darling (AD) and Zhang (ZK, ZA and ZC) 
tests applied to the unbinned data. We also include chi-square tests with equal bin sizes (Equal
Size) and equal bin probabilities (Equal Prob) and the number of bins
found with the often used formula \(k=1+\log_2(n)\). A binning often
done in real life by practitioners uses the data as it would be used to
draw a histogram, that is with a fairly large number of essentially equal
size bins. In the examples below we start with 50 bins but may have a
little less after combining them to achieve $E>5$. This binning is denoted Histogram.
Finally we include our new test, which is always highlighted in the
graphs to make it easier to compare to all others and is called RG.

There are of course many other tests one could use for comparison. There is however none that
is generally much superior to those included here. Moreover, among non-statisticians KS, and to already a
lesser extend AD, are the most commonly used tests because they are included in most software programs.

\subsection{Normal(0, 1) vs t(df)}

We have \(H_0: F=N(0,1)\) vs \(H_a: F=t(df)\). We use a sample size
of \(n=1000\) and \(B=10000\) simulation runs. This will be the case for
all future simulations as well, unless stated otherwise.

\begin{figure}
  \includegraphics[width=\linewidth]{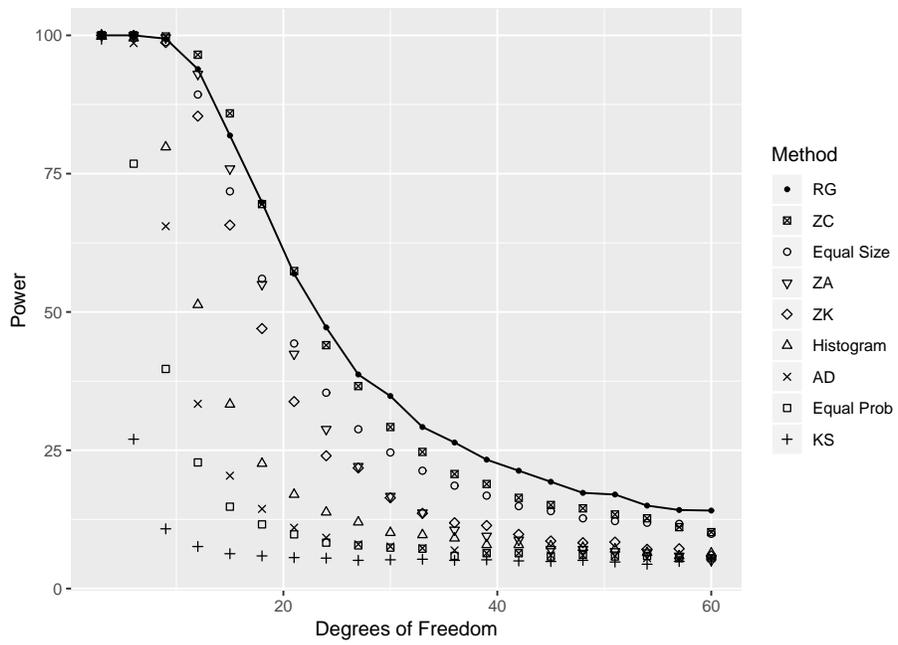}
  \caption{Powers of tests for $H_0:F_0=N(0,1)$ vs $H_a: F_1=t(df)$. ZC is best, with RG second.} 
 \label{figure4}
\end{figure}

The power curves are shown in figure~\ref{figure4}. In the legend the methods are sorted by their mean power, so we see that RG is best and KS is worst. The plotting symbols (a dot for RG) stay the same for all graphs.

The RG test uses 3 bins, \(\kappa=0.5\) and Pearson's formula regardless of sample size.

The mean of the powers over the 20 parameter values are, in order: RG: 46\%. ZC: 44\%,  Equal Size: 40\%, ZA: 36\%, ZK: 34\%, Histogram: 24\%, AD: 21\%, Equal Prob: 18\% and KS: 11\%, so RG is best.

In our studies (not shown here) we always also found the powers of the tests directly and verified that
the combination of $k$, $\kappa$ and test statistic that maximized $M_{k, \kappa, TS}$ 
also had the highest power, or very close to it.

\subsection{Normal vs t(df)}

Here we have \(H_0: F=\mathscr{N}\) vs \(H_a: F=t(df)\), so now the mean and the
standard deviation are estimated. For the KS, AD  and Zhang tests maximum
likelihood estimation is used, and the null distribution is found via
simulation.

\begin{figure}
  \includegraphics[width=\linewidth]{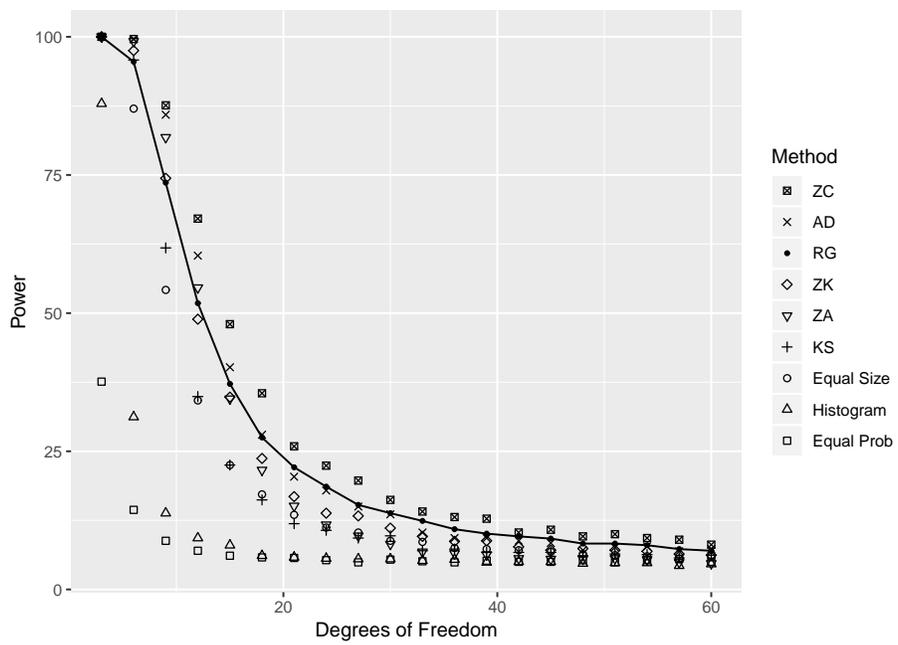}
  \caption{Powers of tests for $H_0:F_0=$Normal vs $H_a: F_1=t(df)$. ZC performs best, with AD second.} 
  \label{figure5}
\end{figure}

The power curves are shown in figure~\ref{figure5}.

RG uses $k=5$ bins, $\kappa=0.5$ and Pearson's formula regardless of sample size. Notice that because two parameters are estimated, four bins is the least possible.

The mean of the powers over the 20 parameter values are, in order: ZC: 31\%, AD: 28\%, RG: 27\%, ZK: 25\%,  ZA: 25\%, KS: 22\%,  Equal Size: 16\%, Histogram: 10\%, Equal Prob: 8\%. RG does is not far behind ZC and about as good as AD.

\subsection{Uniform[0,1] vs Linear}

Next we have \(H_0: F=U[0,1]\) vs \(H_a: F=\text{Linear}(s)\), where the slope changes from 0 (that is the null hypothesis is true) to 0.3.

\begin{figure}
  \includegraphics[width=\linewidth]{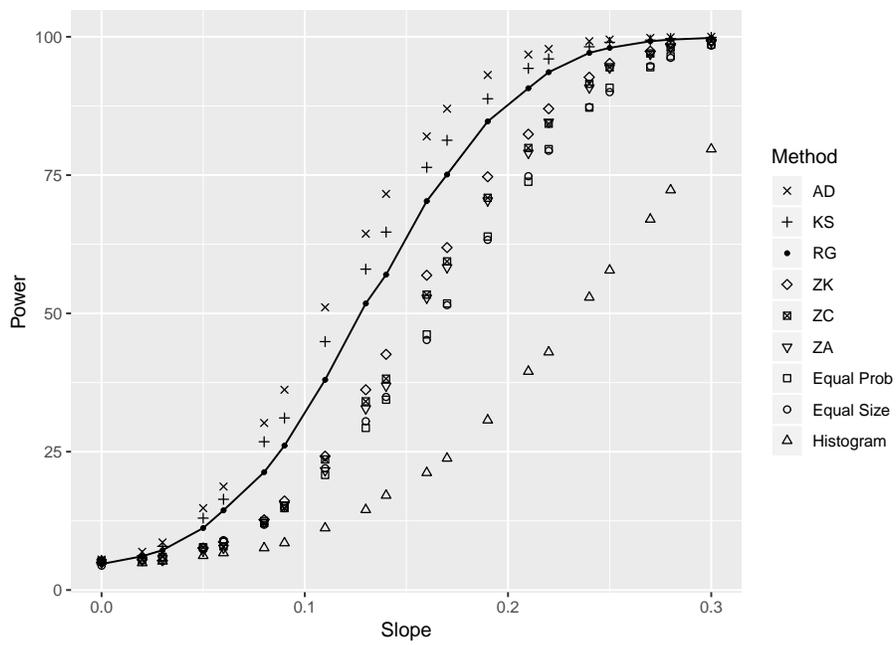} 
  \caption{Powers of tests for $H_0:F_0=U[0,1]$ vs $H_a: F_1=$Linear(slope). AD test performs best, with KS second and RG third.} 
  \label{figure6}
\end{figure}

The power curves are shown in figure~\ref{figure6}.

RG uses 2 or 3 equal probability bins and the Neyman Modified formula regardless of sample size.

The mean of the powers are: AD: 63\%, KS: 60\%, RG: 57\%, ZK: 51\%,  ZA: 49\%, ZC: 49\%, Equal Size: 46\%, Equal Prob: 46\% and Histogram: 29\%. Here the often used histogram binning does especially poorly.

\subsection{Exponential vs Exponential with normal bump}

Here the null hypothesis specifies an exponential with an unknown rate and the alternative is
a mixture of $90\%$ exponential rate 1 and $10\%$ normal mean 1.5 and the
standard deviation varies. The normal is truncated to $[0, \infty)$ and
the rate of the exponential is estimated. This is a case often encountered in high energy physics,
 were the normal bump would indicate the presence of a signal, for example a new particle.

\begin{figure}
  \includegraphics[width=\linewidth]{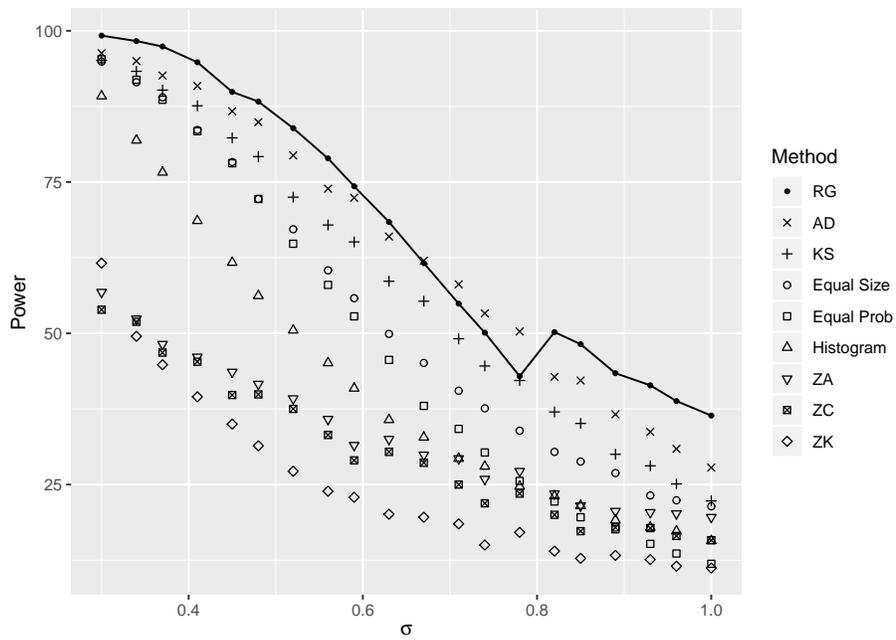}
  \caption{Powers of tests for $H_0:F_0=Exp(1)$ vs exponential with a normal bump. RG performs best for most cases of the parameter, with AD best in a few.} 
  \label{figure7}
\end{figure}

The power curves are shown in figure~\ref{figure7}.

RG uses four bins with the \(\kappa\) depending on the alternative as well as
the sample size. Again Neyman Modified is best.

The mean of the powers are: RG: 67\%, AD: 64\%, KS: 58\%, Equal Size: 53\%, Equal Prob: 48\%, Histogram: 42\%, ZA: 33\%, ZC: 31\% and ZK: 25\%.

The power curve of RG has a curious jump at around $\sigma=0.75$, which is due to the fact
that here $\kappa$ changes from $0.25$ to $0.5$. Choosing
a large number of $\kappa$ values would make this jump disappear. 

This simulation shows that while the Zhang tests generally do very well, sometimes they perform poorly. 

\subsection{Uniform[0,1] versus Beta(1, q)} 

\begin{figure}
  \includegraphics[width=\linewidth]{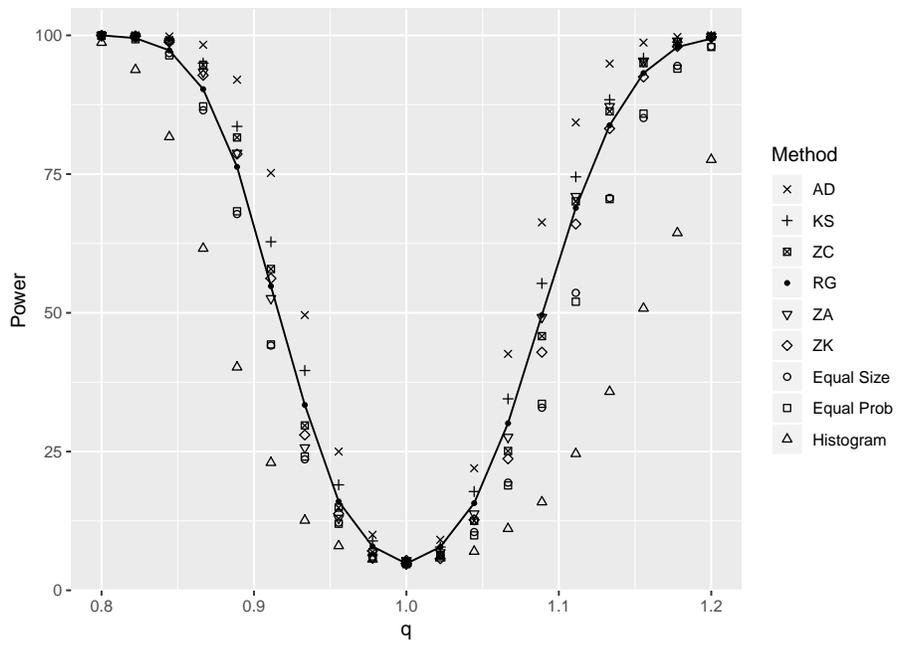}
  \caption{Powers of tests for $H_0:F_0=U[0,1]$ vs Beta(1, q). ZA is best, with AD second and RG third.} 
  \label{figure8}
\end{figure}

The power curves are shown in figure~\ref{figure8}.

RG uses three equal probability bins and Neyman Modifed. 

The mean of the powers are:  AD: 64\%, KS: 60\%, RG: 57\%, ZC: 57\%, ZA: 56\%, ZK: 55\%, Equal Prob: 51\%, Equal Size: 51\% and Histogram: 36\%.

\subsection{Uniform[0,1] versus Beta(q, q)}

\begin{figure}
  \includegraphics[width=\linewidth]{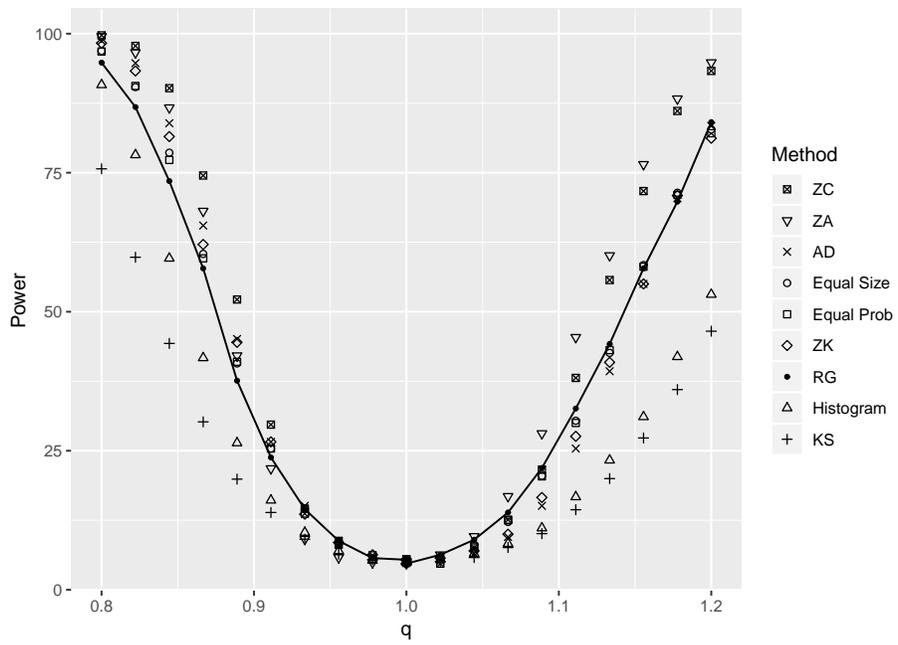}
  \caption{Powers of tests for $H_0:F_0=U[0,1]$ vs Beta(q, q). ZC is best, with RG second.} 
  \label{figure9}
\end{figure}

The power curves are shown in figure~\ref{figure9}.

RG uses six equal probability bins and Neyman Modifed. 

The mean of the powers are: ZA: 44\%, ZC: 44\%, RG: 38\%, Equal Prob: 38\%, Equal Size: 38\%, AD: 38\%, ZK: 38\%, Histogram: 27\% and KS: 22\%.

\subsection{$N(r,\sqrt{r})$ versus Gamma(r, 1)}

\begin{figure}
  \includegraphics[width=\linewidth]{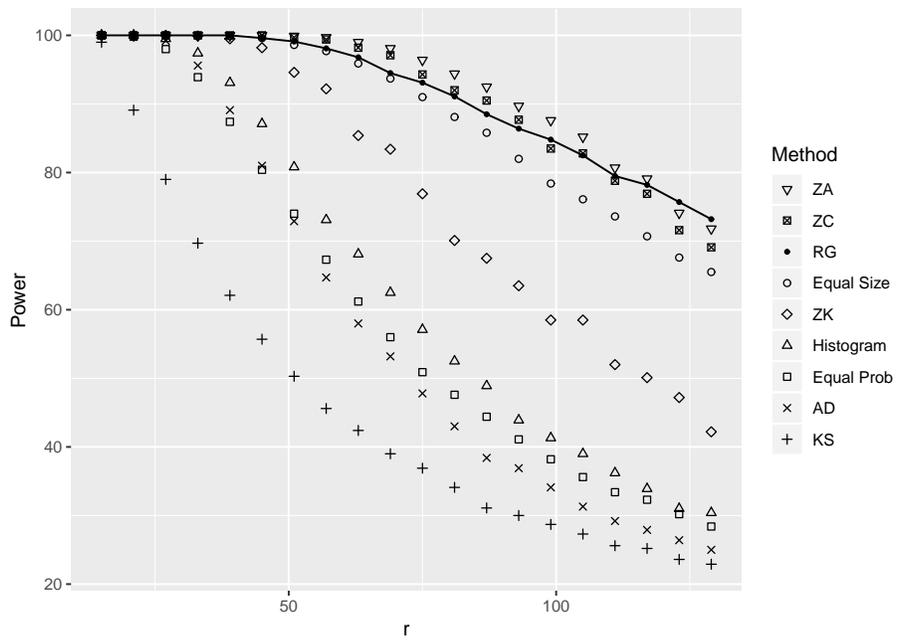}
  \caption{Powers of tests for $H_0:F_0=N(r,\sqrt{r})$ vs Gamma(r, 1). RG is best, with ZC second.} 
  \label{figure10}
\end{figure}

The power curves are shown in figure~\ref{figure10}.

RG uses three equal size bins and Neyman Modifed. 

The mean of the powers are: ZA: 92\%, RG: 91\%, ZC: 91\%, Equal Size: 88\%, ZK: 77\%, Histogram: 64\%, Equal Prob: 60\%, AD: 58\% and KS: 46\%.

This is of course a very small number of simulation studies. However, in the case of 
the goodness-of-fit problem no simulation study can ever be exhaustive. The case studies discussed 
above include a total of 140 power calculations, with powers ranging from the nominal 5\% type I 
error probability to 100\$. The mean power for all cases of the RG method is 54.7\%, which is higher than any other. Those are: ZC 49.5\%, Equal Size 48.2\%, AD 48.0\%, ZA 47.8\%, ZK 43.7\%, KS 39.9\%, Equal Prob 38.4\% and Histogram 33.6\%. That RG has the highest overall power  is due to the fact that it never performs especially badly.

\section{Computational issues}

All the calculations and simulations discussed in this paper were done
using R. An R library with all the routines as well as an RMarkdown file
with the routines to do all the simulations discussed here is available
at https://github.com/wolfgangrolke/chisqalt.

We also created a R Shiny app running online at https://drrolke.shinyapps.io/RGtest/
 that allows the user to upload their data and run the test without knowledge of R. It also allows the 
user to upload a C++ routine for more complicated cases.

\section{Conclusions}

We have presented a new method for binning continuous data for a
chi-square test. Our method uses the idea of a perfect data set to
quickly search through a large number of binnings and finds one which
can be expected to have the highest power from among the included binnings 
against a specified alternative. Our simulation studies show that this method is quite competitive with and sometimes better than either the Kolmogorov-Smirnov, the Anderson Darling or the Zhang 
tests.

Our results also are of interest if there is no alternative hypothesis.
Unlike most published results we find that a small number of bins is
generally best, regardless of the sample size. Certainly the practice in
many fields to draw a histogram with many bins, and then apply the
chi-square test using the same binning leads to a badly underpowered
test.

\section{Figures}

\clearpage
\newpage

\clearpage 

\section{References}

Berkson J. (1980) \emph{Minimum chi-square, not maximum likelihood.}, Ann. Math. Stat. 8(3): 457-487.

Bickel P.J., Doksum K.A. (2015) \emph{Mathematical Statistics Vol 1 and 2}
CRC Press

Birnbaum Z.W. (1952)  \emph{Numerical Tabulation of the Distribution of
Kolmogorov's Statistic for Finite Sample Size.}, JASA 47: 425-441.

Bogdan M. (1995) \emph{Data Driven Version of Pearson's Chi-Square Test for
Uniformity.} Journal of Statistical Computation and Simulation 52:217-237.

Casella G., Berger R. (2002) \emph{Statistical Inference},
Duxbury Advanced Series in Statistics and Decision Sciences. Thomson
Learning.

Cressie N., Read T.R.C (1989) \emph{Pearson's X2 and the Loglikelihood Ratio Statistic G: A
Comparative Review}, International Statistical Review 57: 19-43

D'Agostini R.B, Stephens M.A. (1986) \emph{Goodness-of-Fit
Techniques}, Statistics: Textbooks and Monographs. Marcel Dekker.

Dahiya R.C., Gurland J. (1973) \emph{How Many Classes in the Pearson
Chi-Square Test?}, Journal of the American Statistical
Association, 68:707-712.

Fisher R.A. (1922) \emph{On the Interpretation of Chi-Square of Contingency
Tables and the Calculation of P.}, Journal of the Royal
Statistical Society 85.

Goodman L.A. (1954) \emph{Kolmogorov-Smirnov Tests for Psychological
Research.}, Psychological Bull 51: 160-168.

Greenwood P.E., Nikulin M.S. (1996) \emph{A Guide to
Chi-Square Testing},  Wiley.

Harrison R.H. (1985) \emph{Choosing the Optimum Number of Classes in the Chi-Square Test for
Arbitrary Power Levels.}, Indian J. Stat. 47(3):319-324

Kallenberg W. (1985) \emph{On Moderate and Large Deviations in Multinomial
Distributions.}, The Annals of Statistics 13:1554--1580.

Kallenberg W., Ooosterhoff J., Schriever B. (1985) \emph{The Number of
Classes in Chi-Squared Goodness-of-Fit Tests.}, Journal of the
American Statistical Association 80:959--968.

Koehler K., Gann F. (1990) \emph{Chi-Squared Goodness-of-Fit Tests: Cell
Selection and Power.}, Communications in Statistics-Simulation 19:1265-1278.

Mann H., Wald A. (1942) \emph{On the Choice of the Number and Width of
Classes for the Chi-Square Test of Goodness of Fit.}, Annals of
Mathematical Statistics 13:306-317.

Massey F.J. (1951) \emph{The Kolmogorov-Smirnov Test for Goodness-of-Fit.},
JASA 46: 68-78.

Mineo A. (1979) \emph{A New Grouping Method for the Right Evaluation of the Chi-
Square Test of Goodness-of-Fit.}, Scand. J. Stat. 6(4):145-153.

Ooosterhoff J. (1985) \emph{The Choice of Cells in Chi-Square Tests.},
Statistica Neerlandia 39:115-128.

Quine M., Robinson J. (1985) \emph{Efficiencies of Chi-Square and
Likelihoodratio Goodness-of-Fit Tests.}, Annals of Statistics 13:
727-742.

Raynor J.C., Thas O., Best D.J., (2012) \emph{Smooth Tests of Goodness
of Fit.}.

Sturges H. (1926) The choice of a class-interval. J. Amer Statist. Association 21:65-66

Thas O. (2010) \emph{Continuous Distributions}, Springer Series in
Statistics. Springer.

Voinov N.B., Nikulin M., (2013) \emph{Chi-Square Goodness of Fit Test With Applications.}, 
Academic Press.

Watson G.S.,  (1958) \emph{On Ch-Square Goodness-of-Fit Tests for Continuous Distributions},
Journal of the RSS (Series B) 20:44-72.

Williams CA.  (1950) \emph{On the Choice of the Number and Width of Classes
for the Chi-Square Test of Goodness of Fit.} JASA 45:77-86.

Zhang J. (2002) \emph{Powerful Goodness-of-Fit Tests based on Likelihood Ratio},
Journal of the RSS (Series B) 64:281-294.

\end{document}